\documentstyle[12pt]{article}
\normalsize

\newcounter{sxn}

\newcounter{axn}

\def\br{\begin{thebibliography}{abc}}
\def\er{\end{thebibliography}}

\begin{document}
\bibliographystyle{unsrt}
\footskip 1.0cm
\thispagestyle{empty}
\begin{flushright}
TPI-MINN-92/53-T\\
NSF-ITP-92-145\\
September 1992\\
\end{flushright}
\vspace*{10mm}
\centerline {\Large STRING DEFECTS IN CONDENSED MATTER SYSTEMS}
\vspace*{3mm}
\centerline {\Large  AS OPTICAL FIBERS}
\vspace*{15mm}
\centerline {\large Ajit Mohan Srivastava \footnote{Present address: Institute
for Theoretical Physics, University of California, Santa Barbara,
California 93106, USA}}
\vspace*{5mm}
\centerline {\it Theoretical Physics Institute, University of Minnesota}
\centerline {\it Minneapolis, Minnesota 55455, USA}

\vspace*{15mm}

\baselineskip=18pt

\centerline {\bf ABSTRACT}
\vspace*{8mm}

 We analyze the core structure  of string defects in various condensed
matter systems, such as  nematic liquid crystals and superfluid
helium, and argue that in certain cases the variation of the
refractive index near the core is such that it can lead to total
internal reflection of light travelling along the string core.
These strings thus behave as optical fibers providing a
qualitatively new approach to optical fibers.
We present a candidate for such a fiber by looking at string
segments in a thin nematic liquid crystal film
on water. We discuss various possibilities for constructing such fibers
as well as possible technological applications.

\newpage

 The study of strings has been a subject of great interest in the context
of the early Universe as well as various condensed matter systems
\cite{mermin,gennes,kleman}. Recently, the formation and dynamical
evolution of string defects has been experimentally studied
in nematic liquid crystals with results
which are in good agreement with theoretical expectations \cite{nlc}.
In this letter we investigate an interesting implication of the
existence of string defects.  We consider
strings in certain transparent condensed matter systems and study the
variation of the refractive index near the core of the string.
We then argue that for certain systems the spatial variation of
the refractive index is such
that it leads to total internal reflection of light travelling along the
string core. These strings therefore behave like optical fibers and can
be used for light transmission.

   First let us briefly  recall the essential properties of an optical
fiber \cite{fiber}. An optical fiber  basically consists of an inner
glass (or transparent plastic) wire with an outer
coating of another transparent material such that the refractive
index of the inner material is larger than that of the outer
coating. This leads to the total internal reflection of any light
ray travelling along the inner wire for not too large incidence
angles. There are also fibers which have many coatings with
successively decreasing refractive indices. The thinnest fibers
are typically 10-100 microns thick.

 Now we consider the structure of string cores in condensed matter systems.
Consider a condensed matter system which has two phases
separated by a phase transition such that the low temperature phase
supports strings. Examples are liquid crystals, superfluid
helium etc. \cite{mermin}. We consider only transparent
systems as we are considering optical fibers. We are investigating
extensions of these ideas to strings in other systems, such as flux tubes
in type II superconductors and cosmic strings, and hope to present it
in a future work.  Consider now the core of the string.
The core of the string will typically consist of the high
temperature phase and as one goes away from the center, the phase will
gradually become the low temperature phase.
In other words, if there is  an order parameter $\phi$ such that
$\phi$ = 0 for large temperatures and $\phi$ = some constant, say $\eta$, for
low temperatures, then $\phi$ = 0 on the axis of the string and $\phi$
gradually rises to the value $\phi = \eta$ as we go away
from the center of the string. The distance over which $\phi$
becomes almost $\eta$ is the radius of the string.

 The existence of such strings is typically associated with a winding
number describing the rotation of the order parameter as one encircles
the string. However, The only aspects of such a string, which are
relevant for our discussion, are the nature and stability of the core of
the string and not the winding of the order parameter field around the core.
We will therefore generally not refer
to the winding number and will loosely refer to the magnitude of the order
parameter as the order parameter itself.

 Now suppose that the refractive index (RI) corresponding to a given
wavelength of light has a value $n_1$ in the low temperature phase
(where $\phi = \eta$) and $n_2$ in the high temperature phase
(where $\phi = 0$). Consider the case when
$n_1 < n_2$. Let us now look at the structure of the core of the
string. As we discussed above, the core of the string consists of
the high temperature phase with $\phi = 0$ and as we go away from the core,
$\phi$ gradually becomes equal to $\eta$ as the low temperature phase is
achieved. This then implies that the refractive index has the larger
value $n_2$ inside the core of the string and as we go away from the core,
the refractive index decreases to the smaller value $n_1$ as $\phi$
rises to the value $\eta$. We note that this is the same
as the configuration of an optical fiber as described above.
Any light (with wavelength for which the refractive index has the
assumed values) travelling along the core will undergo total internal
reflection for not too large incidence angles. In fact, from
any physical considerations one will expect that
RI will monotonically decrease from $n_2$ to $n_1$ as $\phi$ increases
from 0 to $\eta$. This string therefore  is like an optical fiber which
consists of very large
number of coatings all with successively decreasing RIs.

 This provides a qualitatively new approach to optical fibers.
Typically optical fibers are manufactured by creating
wires and then coating them; this process limiting the thinness of the fiber
due to technological limitations.
Strings, on the other hand, are typically
of microscopic thickness and for  suitable systems may provide
the thinnest possible optical fibers for a given wavelength of light.
Another interesting property of these string fibers is due to the fact
that typically the core thickness of a string is temperature dependent and
can grow by large factors as one approaches the phase transition temperature.
These fibers will then be such that one could control their thicknesses
by simply changing the temperature so that they are suitable for different
wavelengths of light or for different image resolution requirements.
In contrast the thickness of conventional fibers is pretty
much fixed. The main problem with these string fibers may
be that it will be difficult to create long lengths of these strings,
thereby limiting their possible use only to small scale devices.
Let us now consider some examples of condensed matter systems
where these string fibers may actually be realized.

 First consider vortices in  superfluid $^4$He. Generally these vortices
have cores of atomic dimensions but at
temperatures very close to the $\lambda$ point the cores can be very large,
of the size of few thousand $\AA$ \cite{he4}.
At these temperatures the core of the vortex will typically consist of normal
fluid with a circulation of superfluid in the outer region.
The refractive index data for superfluid $^4$He shows \cite{rihe} that the
value of ($n-1$), where $n$ is the refractive index, at temperature
$T = 1.6K$ is equal to 0.028488 and ($n-1$) monotonically rises to a value of
0.028668 at the $\lambda$ point, T$_\lambda$ = 2.17$K$. The wavelength of
the light used in \cite{rihe} was 5462.27 $\AA$.
As the temperature is increased
above 2.17$K$ $(n-1)$ starts decreasing, though this has no consequence
for us. The important thing for us is that RI is smaller in the superfluid
phase than in the normal phase for temperatures below T$_\lambda$. Therefore,
in the temperature regime where the vortex core is large and consists
of normal fluid the RI should decrease as one goes away from
the core of the vortex. From our earlier discussion we then  see that
these vortices should behave like optical fibers. There are certain points
one has to be careful about. For example the refractive index data refers to
the change in RI at different temperatures which does not necessarily mean
that the RI for normal fluid will be larger than that of the superfluid
when both are at the same temperature (as in a vortex configuration).
These issues will be discussed further in a paper under preparation
\cite{all} where we will also discuss the case of vortices in  superfluid
$^3$He which typically have very large cores with richer structure.

 Now let us consider the example of strings in nematic liquid crystals (NLCs).
NLCs consist of rod like molecules which tend
to locally align in the low temperature $nematic$ phase. In the high
temperature $isotropic$ phase the directions of these molecules are
completely random. The order parameter for isotropic-nematic phase transition
can be taken to be the $director$ $\bf D$ giving the average
orientation (direction and magnitude)
of these  molecules in a small region. In the nematic phase $\bf D$ will
have some fixed magnitude and in the isotropic phase random directions
of the molecules will average out to give $\bf D$ = 0.
In the nematic phase there
are various kinds of topological defects, strings being one of them.
There are strings of strength $S$ where the director $\bf D$
rotates by $2\pi S$ around the core. For half integral $S$ the core has to
be in the isotropic phase but for integral values of $S$ the core may be
in the nematic phase by escape into the third dimension \cite{gennes,kleman}.
The thickness of the isotropic core of the string is of the order $100 \AA$,
though the thickness of the string itself (characterized by the distance from
the core where $\bf D$ becomes significantly non-zero) can be much larger.

 We first describe an experiment where we have observed segments of
strings which seem to behave like optical fibers.
We used the nematic liquid crystal K15 (4-cyano-$4'$-n-pentylbiphenyl)
manufactured by BDH Chemicals, Ltd.. The transition temperature for K15
is 35.3$^0$C. We prepared thin films of nematic
liquid crystal on water and studied them through a phase contrast microscope
(Olympus, Inc. model BH) with a 10X objective. The film was illuminated from
underneath by a standard light source. The microscope was connected to a
monochrome television camera and the image was recorded by a standard
videocassette recorder. The temperature of water was chosen so that the
NLC film is in the nematic phase.

As the NLC film formed on water
surface, we saw bright dots of light which were the ends of string segments.
As the strings shrank, these bright dots came together and
annihilated. When most of the strings disappeared, a slight shaking of
the water surface again created many string segments.
Fig. 1 shows  pictures where one sees strings ending in bright dots.
The strings are always much fainter than the ends which are
typically very bright.

These pictures are very suggestive that these strings may be
trapping light which is entering the string from underneath and then
(due to the curvature of the string) exiting
from the ends of the string. Though we would like to emphasize that
this should only be taken as a suggestion in the sense that there may be
other explanations for the bright ends. For example, it is possible
that the ends of the string segments scatter light in a different way,
especially due to the deformation of the NLC surface near the string ends,
(see \cite{gennes}), and that makes it bright.
Therefore, for nematic liquid crystals, the
following discussion is to be taken as suggesting strong possibility
and our pictures as possible candidates for optical fibers.

It is known that in the nematic phase there are two RIs depending
on the angle between the director $\bf D$  (the order parameter) and
the electric vector $\bf E$ of the light \cite{hndbk}.
If $\bf D$ and $\bf E$ are parallel (extra-ordinary ray) then RI = $r_e$
and when $\bf D$ is  perpendicular  to $\bf E$ (ordinary ray) then RI = $r_o$
such that $r_e > r_o$. In the isotropic phase when $\bf D$ = 0, RI = $r_i$
such that $r_o < r_i < r_e$. For example, for NLC
ethyl p-(p-ethoxybenzyl-ideneamino)-$\alpha$-methylcinnamate, the values of
RI near the transition temperature $T_c \simeq 115^0C$ are, $r_e \simeq 1.75$,
$r_o \simeq 1.55$ and $r_i \simeq 1.62$, see \cite{hndbk} .

Consider now a cylindrical configuration of a string
of strength $S$ = 1 along the z axis where $\bf D$ lies in the x-y plane and
rotates by full $2\pi$
as we go around the string. This string is topologically unstable
in the sense that an isolated string of $S$ = 1 can be continuously deformed
so that the director $\bf D$ points in the same direction everywhere. However,
this deformation may not be possible if the NLC is in some container where
$\bf D$ typically assumes fixed orientation at the boundaries. For example,
for liquid crystals in thin glass tubes with certain types of coatings
in the inner wall of the tube, $\bf D$ becomes normal to the wall and forms
strings of strength $S$ = 1 (where $\bf D$ rotates by $2\pi$) \cite{tube}.
There is a barrier for breaking this string inside the glass tube.
Further assume that, as we approach the center of the cylinder,
a core of isotropic phase is achieved with $\bf D$ = 0.
Note that in this case $\bf D$  does not have to become zero as it can
start pointing along the length of the string. However, it is known that
there are many situations where it is energetically more favorable to have
D = 0 in the middle instead of having nematic phase throughout (see
\cite{tube}). This is the most suitable candidate for an optical fiber
for a NLC.

 Let us assume that unpolarized light is travelling along the core of
such a string. Now suppose that along any direction away from the string
axis where the  director $\bf D$ becomes non-zero, $\bf D$ points in a
certain direction. Then 50\% of light which has
$\bf E$ parallel to $\bf D$ in that region,
will get out of the string rather quickly as for that portion of light
the string acts as an anti optical fiber in the sense that the RI
in the core (where the liquid crystal is in the isotropic phase) is smaller
than the RI away from the core (where the liquid crystal is in the nematic
phase).

  However, for the remaining 50\% of light $\bf E$ will be normal to
$\bf D$ and
this light will undergo total internal reflection at that boundary.
In fact when it bounces off the opposite wall, it will again undergo
total internal reflection as $\bf D$ at opposite points are parallel.
Thus, as long as the reflections remain in a plane, 50\% of
light will undergo total internal reflection at all walls of the
string and for that much of the light the string is like an optical fiber.
Note that it is important for this that as we go along the string
axis, $\bf D$ does not twist. Twists in $\bf D$ will be generally suppressed,
especially at low temperatures, since they cost energy.
As for the elementary string of strength 1/2,
the problem in repeating the above argument is that as the light
bounces on opposite walls it will not find $\bf D$ with the same relative
angle so there will be some light loss. (The string segments we see in NLC
films  on water are probably of this type, though $S$ = 1 strings are also
possible due to special boundary conditions at the NLC-air interface.)
Fixed boundary conditions or applied external electric field
may also give a suitable distribution of $\bf D$ for this case. We will
discuss these points in more detail in \cite{all}.

  The above examples show that there are numerous examples where the
fundamental string defects in condensed matter systems have the
property of trapping light and behave like optical fibers. One needs to
investigate other condensed matter systems to look for such strings.
In the following
we now consider possible ways of creating such fibers. We also comment on
the possibilities of technological applications, though at present these ideas
are at very elementary level.

 As we mentioned above, for NLCs the construction of strength 1 string
seems rather simple. One simply takes a glass tube with appropriate
coating on the inner wall and fills it with NLC. This creates a string
of strength 1 in the tube which under suitable conditions will have isotropic
core \cite{tube} and from the above discussion should behave like
an optical fiber. From the above discussion it should be clear that for
this case it may be necessary to constrain any bending of the tube (and the
trajectory of the light) to remain in a plane to avoid the loss of light
being larger than 50\%.

 For $^4$He, the creation of vortices is even simpler. Imagine that a
cylindrical tube is filled with superfluid helium and this tube is rotated
about its cylindrical axis (initially we may assume this to be a straight
cylinder, though this tube can be bent later and actually even the rotation
can be achieved while the tube is already bent).
This process is well known to create a bundle of vortices
spreading from one end to another  end
of the cylinder \cite{he4}. This bundle of vortices
is completely topologically stable and it may be possible to use
it for image transmission (or for vortex imaging itself).
One has to be careful though as
there may be interchanges in the relative positions of the ends
of the vortices which may scramble the image. However, these strings
repel each other and in steady ground state their ends form a well defined
pattern (such as a triangular lattice), see \cite{he4}. One will expect
that each vortex will remain parallel to the local axis of the cylinder,
this being the shortest
length for the vortex and energetically preferred. This suggests that
there will not be any scrambling of image if one does not keep
moving the cylinder and lets it settle to its lowest energy state.

The important point about such an image transmitting device being that
all these fibers are so thin that they are typically of the order of
the wavelength of the light (or smaller, so one has the worry about
diffraction effects and one may use very short wavelengths)
which is anyway absolute limit to the resolution of an image.
Note that typical size here can be about 1/10 - 1/100 of the size of
conventional fibers, implying about 100- 10,000 times
more information in the image. Note that in this case even the number of
these vortex fibers can be easily manipulated by changing the angular
speed of the cylinder, higher speeds producing larger number of vortices.

 We conclude by emphasizing again the novel aspects of our approach to
optical fibers. Here one just looks for appropriate transparent condensed
matter   systems  which support string like topological excitations
and finds whether the refractive index has correct properties. These strings
then can behave like optical fibers. As these strings are of microscopic
thickness and typically may not be very long, presumably they can find
applications only in small devices.

I am extremely grateful to Edward Lipson and his biophysics group at
Syracuse University for very generously providing access to their optical
equipment and instruction in its use. I am also very  grateful to
Mark Bowick, L. Chandar and Eric Schiff for giving me the opportunity
to work with liquid crystals and for useful discussions. I thank
A.P. Balachandran,  Chandan Dasgupta, Carl Rosenzweig, Michael Stone,
Shikha Varma and William Zimmermann for  useful suggestions.
This research was supported by the US Department of Energy under contract
number DE-AC02-83ER40105, by the Theoretical Physics Institute at the
University of Minnesota and by the National Science Foundation under
Grant No. PHY89-04035, .

\baselineskip=14pt

\baselineskip=18pt

\vskip .3in
\noindent {\Large {\bf Figure Caption}}
\vskip .1in

 Figure 1 : These pictures show bright dots which are the
ends of  string segments. Strings are much fainter than their ends which
appear bright. (a) shows one end of a string. (b) shows a string
segment with both ends visible. (c) shows two string segments with their
ends almost touching each other. The scale of the pictures is shown in (d).

\end{document}